\documentclass[aps,preprint,showpacs,preprintnumbers,amsmath,amssymb]{revtex4}
\usepackage{amsmath,mathrsfs,amsbsy,color,graphicx,bm,amsthm,amsfonts}
\usepackage{units}
\usepackage{bbm}
\usepackage{times}
\usepackage{dcolumn}
\usepackage{amssymb,epsfig}
\newcommand{\udots}{\mathinner{\mskip1mu\raise1pt\vbox{\kern7pt\hbox{.}}
\mskip2mu\raise4pt\hbox{.}\mskip2mu\raise7pt\hbox{.}\mskip1mu}}
\begin{document}
\title{Robustness of network quantum steering under Hawking effect}
\author{ Wen-Mei Li$^1$, Jianbo Lu$^1$\footnote{Email: lvjianbo819@163.com}, Shu-Min Wu$^1$\footnote{
Email: smwu@lnnu.edu.cn}}
\affiliation{$^1$ Department of Physics, Liaoning Normal University, Dalian 116029, China}

% \baselineskip=0.65 cm

%\vspace*{0.2cm}
\begin{abstract}
Network quantum steering provides a framework for certifying quantum correlations in networks with independent sources and partially trusted measurement devices. We investigate how its certification in Schwarzschild spacetime depends on the distribution of nodes affected by Hawking effects and on the initial entanglement. We consider a three-node linear network supplied by two independent sources of entangled massless Dirac-field qubits and compare configurations in which one or both end nodes hover near the event horizon. After tracing over the inaccessible interior modes, we derive analytical expressions for a network steering functional and examine the violation of its associated inequality. Hawking effects suppress the violation, with the double-terminal configuration exhibiting lower robustness than the single-terminal configuration. Moreover, the initial-state parameter yielding the largest violation shifts away from the maximally entangled value as the Hawking temperature increases. In the double-terminal configuration, nonmaximally entangled inputs can retain a violation at temperatures for which maximally entangled inputs are no longer certified by the same criterion. These results identify node configuration and initial-state preparation as joint determinants of certifiable network steering in curved spacetime. They also suggest a practical advantage for platforms where nonmaximally entangled states are easier to prepare, as such inputs can outperform maximally entangled states in preserving certifiable network steering under Hawking effects.
\end{abstract}

\vspace*{0.5cm}
\maketitle

\section{Introduction}
Quantum steering, introduced by Schr\"{o}dinger in the context of the Einstein-Podolsky-Rosen paradox \cite{M18,M19}, enables entanglement certification with untrusted measurement devices \cite{M20,M24,M25} and occupies an intermediate position between entanglement \cite{M21,M22} and Bell nonlocality \cite{M23}. Quantum networks extend the study of these correlations to multiple nodes connected by independent sources, providing a setting for distributed quantum communication \cite{M43,M44} and the characterization of network topologies \cite{M45}. Source independence imposes constraints on network correlations, motivating theoretical studies of bilocal and more general configurations \cite{M46,M47,M48,M49}, further theoretical and experimental advances \cite{M50,M51,M52}, and a systematic account of network nonlocality \cite{M53}. When a network contains both trusted and untrusted measurement devices, network quantum steering tests the compatibility of conditional states with a network local hidden state model \cite{M54}. Related progress includes multipartite steering certification \cite{M55,M56,M57}, bilocal inequalities involving elegant joint measurements \cite{M58}, further steering detection methods \cite{M60,M61,M62}, and randomness certification based on multipartite steering \cite{M63}. These developments motivate investigating the robustness of network steering under physical noise.

Relativistic quantum information studies how quantum information is encoded, processed, and transmitted in settings where relativistic motion and spacetime geometry affect quantum states and their correlations \cite{SDF1,SDF2,SDF3,SDF4,SDF5,SDF6,SDF7,SDF8,SDF9,SDF10,SDF11,SDF12,SDF13,SDF14,SDF15,SDF16,SDF17,SDF18,SDF19,SDF20,AAASDF59,SDF21,SDF22,SDF23,SDF24,SDF25,SDF26,SDF27,SDF28,SDF29,SDF30,SDF31,SDF32,SDF33,SDF34,SDF35,SDF36,SDF37,SDF38,SDF39,SDF40,SDF41,SDF42,SDF43,SD1,QGL1,QGL2,SD2,QGL3,QGL4,SD3,QGL5,QGL6,QGL7,QGL8,QGL9,QGL10,SDF45,SDF46,SDF47,SDF48,SDF49,SDF50, SDF51,SDF52,SDF53}. A central concern is how the information accessible to different observers determines the quantum resources available for information processing \cite{AAASDF59,SDF21,SDF22}. Within this framework, the robustness of network quantum steering in curved spacetime deserves further investigation, since its certification depends on both the correlations shared between nodes and the independence of the underlying sources. In a black hole background \cite{SDF1,SDF2,SDF3,SDF4,SDF5,SDF6}, tracing over field modes inaccessible beyond the event horizon modifies the accessible states distributed by these sources, raising the question of how Hawking effects influence the certification of network steering. The network structure introduces an additional consideration: the degradation may depend on the distribution of nodes exposed to Hawking effects. It is therefore important to compare configurations in which one or both end nodes are affected and to determine how this difference changes the persistence of a steering-inequality violation. A related question concerns the role of initial entanglement. Although maximally entangled states are natural candidates for distributing strong quantum correlations, they need not remain optimal after the accessible states undergo Hawking-induced degradation. This question also has practical relevance: on platforms where nonmaximally entangled states are easier to prepare, greater robustness of these inputs would combine reduced preparation demands with improved performance in relativistic quantum information tasks based on network steering. Whether a maximally entangled input yields the strongest violation of a network steering inequality, or whether nonmaximally entangled states can offer greater robustness, thus requires explicit examination. These questions motivate a comparative study of different node configurations and initial-state parameters in a Schwarzschild black hole background.

In this work, we investigate network quantum steering in a three-node linear network formed by two independent sources of entangled massless Dirac-field qubits in a Schwarzschild black hole background. We compare two configurations: one in which only Charlie hovers near the event horizon, while Alice and Bob remain in the asymptotically flat region, and another in which both Alice and Charlie hover near the horizon, while Bob remains far from the black hole. For each configuration, we obtain the accessible states by tracing over the field modes inside the horizon and derive analytical expressions for the network steering functional based on the inequality proposed by Li et al.~\cite{MM}. We examine how its violation depends on the Hawking temperature $T$ and the initial-state parameter $\theta$. The results show that Hawking effects suppress the inequality violation, with the configuration involving two affected end nodes exhibiting weaker robustness than that involving a single affected end node. We further analyze the temperature dependence of the initial-state parameter that yields the largest violation, addressing whether maximally entangled inputs remain optimal for this certification task. Together, these results elucidate how node configuration and initial-state preparation influence the certification of network quantum steering in curved spacetime.

The remainder of this paper is organized as follows. Section~\ref{sec:dirac} reviews the quantization of the Dirac field in Schwarzschild spacetime. Section~\ref{sec:network-steering} introduces the network steering criterion used in our analysis. Section~\ref{sec:results} presents the two node configurations and examines the dependence of the steering-inequality violation on the Hawking temperature and initial-state parameter in curved spacetime. Section~\ref{sec:conclusions} summarizes our findings.

\section{Quantization of the Dirac field in Schwarzschild spacetime\label{sec:dirac}}

We first specify the field-mode description used to model the Hawking effect on the network states. In units $G=c=\hbar=k_{B}=1$, the Schwarzschild metric is
\begin{equation}\label{S1}
ds^{2}=-f(r)dt^{2}+f(r)^{-1}dr^{2}+r^{2}(d\vartheta^{2}+\sin^{2}\vartheta\,d\varphi^{2}),
\qquad f(r)=1-\frac{2M}{r},
\end{equation}
where $M$ is the black hole mass, $r$ is the areal radial coordinate, and the event horizon lies at $r_{H}=2M$. The angular coordinate $\vartheta$ is distinguished from the initial-state parameter $\theta$ used below. The surface gravity is $\kappa=1/(4M)$, giving the Hawking temperature $T=\kappa/(2\pi)=1/(8\pi M)$.

To describe modes across the horizon, we introduce the tortoise coordinate $r_{*}=r+2M\ln|r/(2M)-1|$ and the null coordinates $u=t-r_{*}$ and $v=t+r_{*}$. Up to constant scale factors, the Kruskal coordinates in the exterior region are
\begin{equation}\label{S0}
U=-e^{-\kappa u},\qquad V=e^{\kappa v}.
\end{equation}
Across the future event horizon, $U$ changes sign, while $V$ remains positive. These coordinates remove the coordinate singularity at $r=2M$ and permit the analytic continuation of the near-horizon modes.

We consider a massless Dirac field $\Phi$, whose covariant equation of motion is
\begin{equation}\label{S2}
\gamma^{a}e_{a}{}^{\mu}(\partial_{\mu}+\Gamma_{\mu})\Phi=0,
\end{equation}
where $e_{a}{}^{\mu}$ is an inverse tetrad, $\gamma^{a}$ are the flat-space Dirac matrices, and $\Gamma_{\mu}$ is the spinor connection \cite{SDF54}. Using a diagonal tetrad adapted to the Schwarzschild coordinates, Eq.~(\ref{S2}) takes the explicit form
\begin{equation}\label{w12}
\begin{aligned}
&-\frac{\gamma_{0}}{\sqrt{f(r)}}\frac{\partial\Phi}{\partial t}
+\gamma_{1}\sqrt{f(r)}\left[\frac{\partial}{\partial r}
+\frac{1}{r}+\frac{M}{2r(r-2M)}\right]\Phi\\
&\quad+\frac{\gamma_{2}}{r}\left(\frac{\partial}{\partial\vartheta}
+\frac{\cot\vartheta}{2}\right)\Phi
+\frac{\gamma_{3}}{r\sin\vartheta}\frac{\partial\Phi}{\partial\varphi}=0,
\end{aligned}
\end{equation}
where $f(r)=1-2M/r$ and $\gamma_{i}$ ($i=0,1,2,3$) denote the Dirac matrices in the local orthonormal frame, with $\gamma_{0}$ denoting the timelike matrix in this component notation \cite{WuZeng2022}. The connection terms account for the radial and angular dependence of the tetrad. In the near-horizon limit, the relevant outgoing modes have the phase dependence
\begin{equation}\label{S3}
\Phi_{\mathbf{k},\mathrm{I}}^{+}\sim\xi_{\mathbf{k}}e^{-i\omega u},
\qquad
\Phi_{\mathbf{k},\mathrm{II}}^{+}\sim\widetilde{\xi}_{\mathbf{k}}e^{i\omega u},
\end{equation}
where $\omega>0$ is the exterior Killing frequency, $\mathbf{k}$ collectively labels the modes, and the spinor amplitudes include the remaining spatial dependence. Here I denotes the exterior and II the black hole interior; these labels should not be confused with the numbering of the four regions in a full Kruskal diagram. The horizon prevents interior signals from reaching exterior observers.

In a basis adapted to the two regions, the field expansion can be written schematically as
\begin{equation}\label{S4}
\Phi=\sum_{\sigma=\mathrm{I},\mathrm{II}}\int d\mathbf{k}\,
\left[a_{\mathbf{k}}^{\sigma}\Phi_{\mathbf{k},\sigma}^{+}
+b_{\mathbf{k}}^{\sigma\dagger}\Phi_{\mathbf{k},\sigma}^{-}\right],
\end{equation}
where $a_{\mathbf{k}}^{\sigma}$ and $b_{\mathbf{k}}^{\sigma}$ annihilate particles and antiparticles, respectively. The superscripts $+$ and $-$ on the modes denote the corresponding particle and antiparticle solutions. The operators obey
\begin{equation*}
\{a_{\mathbf{k}}^{\sigma},a_{\mathbf{k}'}^{\sigma'\dagger}\}
=\{b_{\mathbf{k}}^{\sigma},b_{\mathbf{k}'}^{\sigma'\dagger}\}
=\delta_{\sigma\sigma'}\delta(\mathbf{k}-\mathbf{k}'),
\end{equation*}
with all other anticommutators vanishing. The associated regional vacuum is annihilated by both particle and antiparticle annihilation operators.

Analytic continuation across the horizon yields Kruskal modes that are regular there \cite{AAAA3}. With a suitable phase convention, two unnormalized particle-mode combinations are \cite{SDF53}
\begin{align}\label{S6}
\Psi_{\mathbf{k},\mathrm{I}}^{+}
&=e^{\omega/(4T)}\Phi_{\mathbf{k},\mathrm{I}}^{+}
-e^{-\omega/(4T)}\Phi_{-\mathbf{k},\mathrm{II}}^{-},\notag\\
\Psi_{\mathbf{k},\mathrm{II}}^{+}
&=e^{\omega/(4T)}\Phi_{\mathbf{k},\mathrm{II}}^{+}
+e^{-\omega/(4T)}\Phi_{-\mathbf{k},\mathrm{I}}^{-}.
\end{align}
Together with the corresponding antiparticle modes, they give the alternative expansion
\begin{equation}\label{S7}
\Phi=\sum_{\sigma=\mathrm{I},\mathrm{II}}\int
\frac{d\mathbf{k}}{\sqrt{2\cosh[\omega/(2T)]}}\,
\left[c_{\mathbf{k}}^{\sigma}\Psi_{\mathbf{k},\sigma}^{+}
+d_{\mathbf{k}}^{\sigma\dagger}\Psi_{\mathbf{k},\sigma}^{-}\right].
\end{equation}
The operators $c_{\mathbf{k}}^{\sigma}$ and $d_{\mathbf{k}}^{\sigma}$ annihilate the Kruskal vacuum. Comparing Eqs.~(\ref{S4}) and (\ref{S7}) gives, for the exterior particle sector,
\begin{align}\label{S8}
c_{\mathbf{k}}^{\mathrm{I}}&=\alpha a_{\mathbf{k}}^{\mathrm{I}}
-\beta b_{-\mathbf{k}}^{\mathrm{II}\dagger},\notag\\
c_{\mathbf{k}}^{\mathrm{I}\dagger}&=\alpha a_{\mathbf{k}}^{\mathrm{I}\dagger}
-\beta b_{-\mathbf{k}}^{\mathrm{II}},
\end{align}
where $\alpha=(1+e^{-\omega/T})^{-1/2}$ and $\beta=(1+e^{\omega/T})^{-1/2}$ satisfy $\alpha^{2}+\beta^{2}=1$.

For the network analysis, we retain one exterior particle mode and its interior antiparticle partner at a fixed frequency, suppressing the remaining mode and spin labels.  With the exterior creation operator ordered before its interior partner, the normalized vacuum in this sector is \cite{SDF1,SDF53}
\begin{equation}\label{S9}
|0\rangle_{K}
=\alpha|0_{\mathbf{k}}\rangle_{\mathrm{I}}^{+}|0_{-\mathbf{k}}\rangle_{\mathrm{II}}^{-}
+\beta|1_{\mathbf{k}}\rangle_{\mathrm{I}}^{+}|1_{-\mathbf{k}}\rangle_{\mathrm{II}}^{-},
\end{equation}
and the corresponding one-particle excitation is
\begin{equation}\label{S10}
|1\rangle_{K}=c_{\mathbf{k}}^{\mathrm{I}\dagger}|0\rangle_{K}
=|1_{\mathbf{k}}\rangle_{\mathrm{I}}^{+}|0_{-\mathbf{k}}\rangle_{\mathrm{II}}^{-}.
\end{equation}
Here the superscripts $+$ and $-$ on the number states distinguish particle and antiparticle sectors. Since an exterior observer has no access to the interior partner, the physically accessible state is obtained by tracing over region II. Eqs.~(\ref{S9}) and (\ref{S10}) therefore specify the mode transformation applied independently to each affected network end node.

\section{Detection of network quantum steering\label{sec:network-steering}}

We consider a tripartite linear network consisting of Alice, Bob, and Charlie. Two independent sources distribute the bipartite states $\rho_{AB}$ and $\rho_{B'C}$, respectively, so that the joint state is $\rho_{AB}\otimes\rho_{B'C}$. Alice and Charlie each hold one subsystem, whereas Bob holds the two subsystems $B$ and $B'$, one from each source. A suitable joint measurement on Bob's subsystems can establish conditional entanglement between the end nodes through entanglement swapping. In the steering scenario considered here, Alice's and Charlie's measurement devices are trusted, while Bob's device is uncharacterized.

Let $\{M_b^{BB'}\}_b$ denote the positive-operator-valued measure describing Bob's joint measurement, with $M_b^{BB'}\geq0$ and $\sum_b M_b^{BB'}=I_{BB'}$. Conditioned on his outcome $b$, Alice and Charlie share the unnormalized state
\begin{eqnarray}\label{S11}
\sigma_b^{AC}=\operatorname{Tr}_{BB'}\left[
\left(I_A\otimes M_b^{BB'}\otimes I_C\right)
\left(\rho_{AB}\otimes\rho_{B'C}\right)\right].
\end{eqnarray}
The probability of this outcome is $p(b)=\operatorname{Tr}\sigma_b^{AC}$, and the normalized conditional state is $\sigma_b^{AC}/p(b)$ for $p(b)>0$. The collection of conditional operators $\{\sigma_b^{AC}\}_b$, referred to as the network assemblage, contains the information required to test network steering using the trusted measurements at the end nodes.

The assemblage admits a network local hidden state (NLHS) model if all outcomes can be described by a common decomposition \cite{M54}
\begin{eqnarray}\label{S12}
\sigma_b^{AC}=\sum_{\lambda,\nu}p(\lambda)p(\nu)
 p(b|\lambda,\nu)\,\sigma_A^{\lambda}\otimes\sigma_C^{\nu}.
\end{eqnarray}
Here $\lambda$ and $\nu$ label the independent hidden variables associated with the two sources, $\sigma_A^{\lambda}$ and $\sigma_C^{\nu}$ are normalized local hidden states, and $p(b|\lambda,\nu)$ is Bob's response function. The factorized distribution $p(\lambda)p(\nu)$ expresses source independence. Network steering is certified when no such model reproduces the assemblage; the decomposition must therefore hold simultaneously for every $b$, rather than separately for each conditional state.

To obtain a directly testable criterion, we employ the network Clauser-Horne-Shimony-Holt-like inequality proposed by Li et al.~\cite{MM}. In the quantum realization, Bob performs a fixed four-outcome elegant joint measurement (EJM). Its outcome is assigned one of the triples $(1,1,1)$, $(1,-1,-1)$, $(-1,1,-1)$, and $(-1,-1,1)$. Writing $b_j$ for the $j$th component of the assigned triple, $B^j$ ($j=1,2,3$) denotes the corresponding recorded variable, represented quantum mechanically by $B^j=\sum_b b_jM_b^{BB'}$. Thus the three variables are extracted from a single joint measurement, rather than from three independent measurement settings.

Alice and Charlie measure the qubit observables $A_x=\hat{\mathbf{u}}_x\cdot\vec{\sigma}$ and $C_y=\hat{\mathbf{v}}_y\cdot\vec{\sigma}$, with $x,y=1,2,3$ and outcomes $a,c=\pm1$. Here $\vec{\sigma}=(\sigma_1,\sigma_2,\sigma_3)$ is the Pauli vector, and the measurement directions form an orthonormal triad at each end node: $\hat{\mathbf{u}}_x\cdot\hat{\mathbf{u}}_{x'}=\delta_{xx'}$ and $\hat{\mathbf{v}}_y\cdot\hat{\mathbf{v}}_{y'}=\delta_{yy'}$. Their collection is denoted by $\mu=\{\hat{\mathbf{u}}_1,\hat{\mathbf{u}}_2,\hat{\mathbf{u}}_3,\hat{\mathbf{v}}_1,\hat{\mathbf{v}}_2,\hat{\mathbf{v}}_3\}$. The correlators are evaluated from the observed statistics as $\langle A_xB^jC_y\rangle=\sum_{a,b,c}ab_jc\,p(a,b,c|x,y)$. Under these conditions, every NLHS model satisfies
\begin{eqnarray}\label{S13}
F\left(\mu,\rho_{AB}\otimes\rho_{B'C}\right)&&=
\sqrt{\sum_{x,y=1}^{3}\left\langle A_x(B^2+B^3)C_y\right\rangle^2}
+\sqrt{\sum_{x,y=1}^{3}\left\langle A_x(B^1-B^2)C_y\right\rangle^2}\notag\\
&&+\sqrt{\sum_{x,y=1}^{3}\left\langle A_x(B^1-B^3)C_y\right\rangle^2}\leq2.
\end{eqnarray}
A violation of Eq.~(\ref{S13}), namely $F>2$, rules out the NLHS model and certifies network quantum steering. By contrast, $F\leq2$ only means that this criterion does not detect steering and does not prove that the assemblage is unsteerable. We use $F$ below as a network steering functional for the specified measurement configuration, rather than as a general measure of network steerability.

\section{Network quantum steering in Schwarzschild spacetime\label{sec:results}}

\begin{figure}
\begin{minipage}[t]{0.5\linewidth}
\centering
\includegraphics[width=\linewidth]{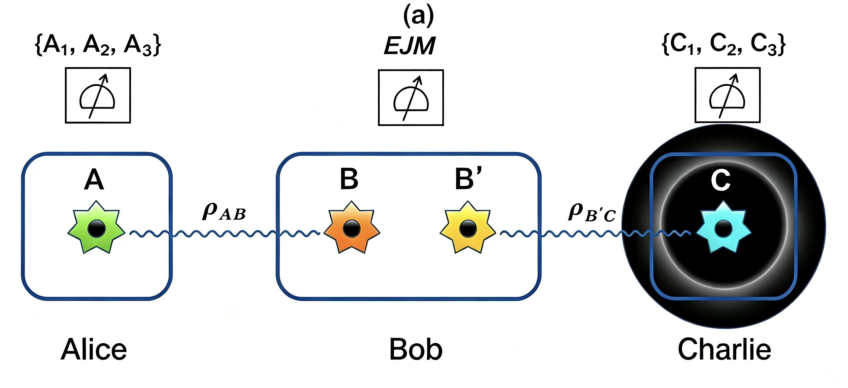}   % 只指定宽度，高度自动
\end{minipage}%
\begin{minipage}[t]{0.5\linewidth}
\centering
\includegraphics[width=\linewidth]{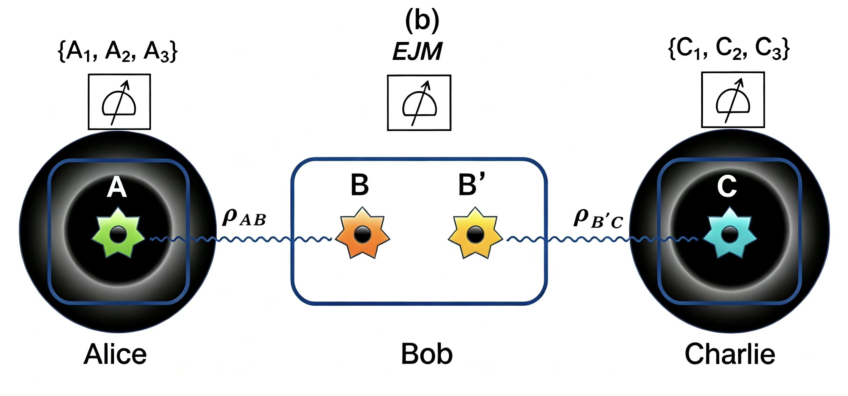}
\end{minipage}%
\caption{Three-node linear network supplied by two independent sources distributing $\rho_{AB}$ and $\rho_{B'C}$. Bob holds subsystems $B$ and $B'$ and performs a joint measurement, while Alice and Charlie perform trusted local measurements. (a) Single-terminal configuration: Alice and Bob remain in the asymptotically flat region, while Charlie hovers near the Schwarzschild event horizon. (b) Double-terminal configuration: Bob remains in the asymptotically flat region, while Alice and Charlie hover near the horizon.}
\label{Fig1}
\end{figure}
The two configurations considered here are shown in Fig.~\ref{Fig1}. In configuration (a), only Charlie hovers near the event horizon, while Alice and Bob remain in the asymptotically flat region. In configuration (b), both Alice and Charlie hover near the horizon, while Bob remains in the asymptotically flat region. Bob performs the same fixed joint measurement in both configurations, and the end-node observables are chosen identically to isolate the effect of the affected nodes.

For the bilocal network shown in Fig.~\ref{Fig1}, the initial resource consists of two independently prepared entangled pairs, shared between $A$ and $B$ and between $B'$ and $C$, respectively. The four-qubit input state is
\begin{equation}\label{S14}
|\Psi_{\mathrm{in}}\rangle_{ABB'C}
=|\phi_{AB}\rangle\otimes|\phi_{B'C}\rangle,
\end{equation}
where
\begin{equation}\label{S15}
|\phi_{AB}\rangle=\cos\theta|0_A0_B\rangle+\sin\theta|1_A1_B\rangle,
\end{equation}
and
\begin{equation}\label{S16}
|\phi_{B'C}\rangle=\cos\theta|0_{B'}0_C\rangle+\sin\theta|1_{B'}1_C\rangle.
\end{equation}
We take the same preparation parameter $0\leq\theta\leq\pi/2$ for both sources. This common parameter specifies their independently prepared states and does not introduce correlations between the sources. At $\theta=\pi/4$, each pair reduces to the Bell state $|\Phi^+\rangle=(|00\rangle+|11\rangle)/\sqrt{2}$, so the input is a product of two Bell states. Other values of $\theta$ describe partially entangled pairs, with product states at $\theta=0$ and $\theta=\pi/2$. The network thus comprises three nodes but four qubit subsystems, since Bob holds both $B$ and $B'$.

For the subsequent calculations, we fix the characterized end-node measurements to the Cartesian Pauli directions,
\begin{equation*}
\mu_0=\{\hat{\mathbf{u}}_1=\hat{\mathbf{v}}_1=(1,0,0),\;
\hat{\mathbf{u}}_2=\hat{\mathbf{v}}_2=(0,1,0),\;
\hat{\mathbf{u}}_3=\hat{\mathbf{v}}_3=(0,0,1)\}.
\end{equation*}
These directions satisfy the orthogonality conditions required by Eq.~(\ref{S13}). We use $\mu=\mu_0$ in both configurations to compare their responses to the Hawking effect for the same end-node measurement settings.
For definiteness, the fixed joint measurement used to obtain the expressions below is the Bell-measurement limit of the generalized EJM, up to local basis rotations \cite{MM}. We choose its three outcome observables as
\begin{equation}\label{eq:bob-observables}
B^1=-\sigma_2\otimes\sigma_3,\qquad
B^2=-\sigma_3\otimes\sigma_1,\qquad
B^3=-\sigma_1\otimes\sigma_2,
\end{equation}
where the tensor factors refer to $B$ and $B'$. These observables commute and satisfy $B^1B^2=B^3$. Their joint eigenvalues are the four outcome triples specified in Sec.~\ref{sec:network-steering}; the corresponding rank-one projectors are $M_b^{BB'}=(I_{BB'}+\sum_{j=1}^{3}b_jB^j)/4$. This specification fixes the measurement entering $F$ and makes the calculation reproducible.

In the single-terminal configuration, we apply Eqs.~(\ref{S9}) and (\ref{S10}) to Charlie's mode in Eq.~(\ref{S16}), while leaving Bob's mode unchanged. The resulting state is
\begin{equation}\label{S17}
\begin{aligned}
|\phi_{B'C_{\mathrm{I}}C_{\mathrm{II}}}\rangle
={}&\alpha\cos\theta|0_{B'}0_{C_{\mathrm{I}}}0_{C_{\mathrm{II}}}\rangle\\
&+\beta\cos\theta|0_{B'}1_{C_{\mathrm{I}}}1_{C_{\mathrm{II}}}\rangle
+\sin\theta|1_{B'}1_{C_{\mathrm{I}}}0_{C_{\mathrm{II}}}\rangle.
\end{aligned}
\end{equation}
The first source state remains pure,
\begin{equation}\label{S18}
\begin{aligned}
\rho_{AB}={}&\cos^2\theta|00\rangle\langle00|+\sin^2\theta|11\rangle\langle11|\\
&+\sin\theta\cos\theta\left(|00\rangle\langle11|+|11\rangle\langle00|\right).
\end{aligned}
\end{equation}
Tracing Eq.~(\ref{S17}) over the inaccessible interior mode $C_{\mathrm{II}}$ gives
\begin{equation}\label{S19}
\begin{aligned}
\rho_{B'C_{\mathrm{I}}}={}&\alpha^2\cos^2\theta|00\rangle\langle00|
+\beta^2\cos^2\theta|01\rangle\langle01|\\
&+\sin^2\theta|11\rangle\langle11|
+\alpha\sin\theta\cos\theta\left(|00\rangle\langle11|+|11\rangle\langle00|\right),
\end{aligned}
\end{equation}
where the basis order is $B'C_{\mathrm{I}}$. For compactness, define $s_\theta=\sin(2\theta)$ and $z_\theta=1-2\beta^2\cos^2\theta$. Substitution into Eq.~(\ref{S13}) yields
\begin{equation}\label{S20}
\begin{aligned}
F_1&\equiv F(\mu_0,\rho_{AB}\otimes\rho_{B'C_{\mathrm{I}}})\\
&=s_\theta\left[\sqrt{z_\theta^2+\alpha^2s_\theta^2}
+\sqrt{\alpha^2+z_\theta^2}+\alpha\sqrt{1+s_\theta^2}\right].
\end{aligned}
\end{equation}

In the double-terminal configuration, the same mode transformation also acts on Alice's subsystem. We assume equal mode frequencies and the same Hawking parameters $\alpha$ and $\beta$ for the two affected end nodes. The first pair then becomes
\begin{equation}\label{S21}
\begin{aligned}
|\phi_{A_{\mathrm{I}}A_{\mathrm{II}}B}\rangle
={}&\alpha\cos\theta|0_{A_{\mathrm{I}}}0_{A_{\mathrm{II}}}0_B\rangle\\
&+\beta\cos\theta|1_{A_{\mathrm{I}}}1_{A_{\mathrm{II}}}0_B\rangle
+\sin\theta|1_{A_{\mathrm{I}}}0_{A_{\mathrm{II}}}1_B\rangle.
\end{aligned}
\end{equation}
Tracing over $A_{\mathrm{II}}$ gives
\begin{equation}\label{S22}
\begin{aligned}
\rho_{A_{\mathrm{I}}B}={}&\alpha^2\cos^2\theta|00\rangle\langle00|
+\beta^2\cos^2\theta|10\rangle\langle10|\\
&+\sin^2\theta|11\rangle\langle11|
+\alpha\sin\theta\cos\theta\left(|00\rangle\langle11|+|11\rangle\langle00|\right),
\end{aligned}
\end{equation}
with basis order $A_{\mathrm{I}}B$. The accessible network state is $\rho_{A_{\mathrm{I}}B}\otimes\rho_{B'C_{\mathrm{I}}}$, for which
\begin{equation}\label{S23}
\begin{aligned}
F_2&\equiv F(\mu_0,\rho_{A_{\mathrm{I}}B}\otimes\rho_{B'C_{\mathrm{I}}})\\
&=2\alpha s_\theta\sqrt{z_\theta^2+\alpha^2s_\theta^2}
+\sqrt{2}\alpha s_\theta z_\theta.
\end{aligned}
\end{equation}
The zero-temperature expressions are understood as $T\to0^+$, for which $\alpha\to1$ and $\beta\to0$. Both configurations then give $F=2s_\theta\sqrt{1+s_\theta^2}+\sqrt{2}s_\theta$, attaining $3\sqrt{2}$ at $\theta=\pi/4$.

\begin{figure}
\begin{minipage}[t]{0.5\linewidth}
\centering
\includegraphics[width=\linewidth]{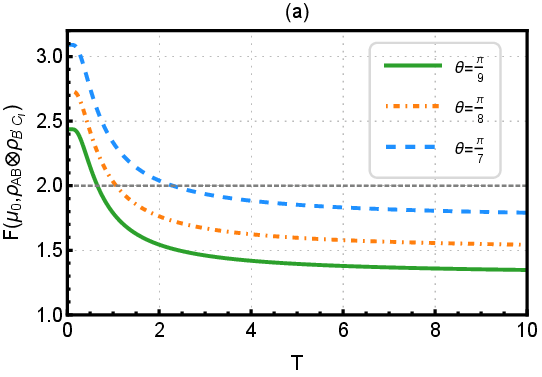}
\end{minipage}%
\begin{minipage}[t]{0.5\linewidth}
\centering
\includegraphics[width=\linewidth]{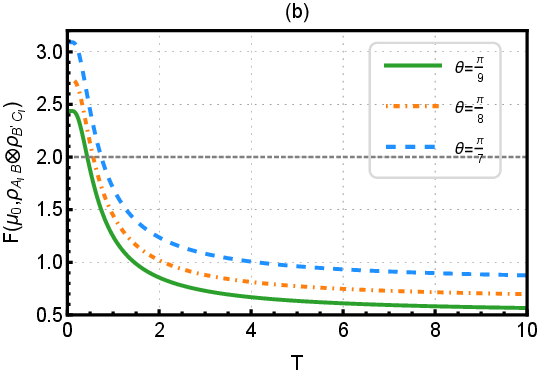}
\end{minipage}%
\caption{Network steering functional $F$ versus Hawking temperature $T$ for $\theta=\pi/9$, $\pi/8$, and $\pi/7$, with $\omega=1$. (a) Single-terminal configuration. (b) Double-terminal configuration. Values above the NLHS bound $F=2$ certify network quantum steering.}
\label{Fig2}
\end{figure}
Fig.~\ref{Fig2} shows the network steering functional $F$, defined in Eq.~(\ref{S13}), as a function of the Hawking temperature $T$ for different initial-state parameters $\theta$, with $\omega=1$. In both configurations, $F$ decreases monotonically with increasing $T$, reflecting the suppression of the network steering inequality violation by Hawking effects. The bound $F=2$ separates the region in which network steering is certified from the region in which the inequality is no longer violated: $F>2$ certifies network steering, whereas $F\leq2$ indicates that steering is not detected by this criterion. For the parameter values displayed, larger $\theta$ produces a larger $F$ and enhances the persistence of the inequality violation against increasing temperature. Comparing Figs.~\ref{Fig2}(a) and \ref{Fig2}(b), the single-terminal configuration retains a larger value of $F$ than the double-terminal configuration under otherwise identical conditions. When both end nodes are affected, tracing over the inaccessible interior modes modifies both source states, leading to a more pronounced reduction of the correlations entering $F$. Consequently, whenever the curves cross the detection bound, the double-terminal configuration reaches it at a lower temperature. This comparison highlights the role of node configuration in preserving certifiable network steering near a black hole.

\begin{figure}
\begin{minipage}[t]{0.5\linewidth}
\centering
\includegraphics[width=\linewidth]{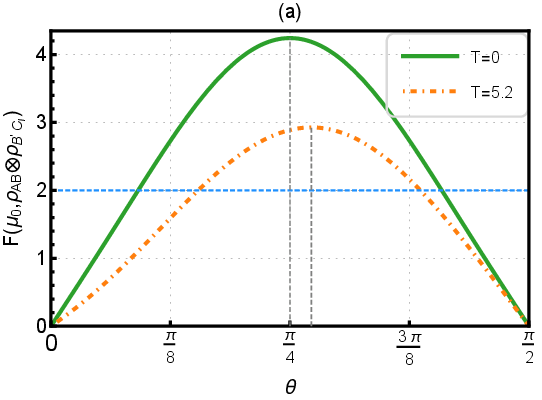}
\end{minipage}%
\begin{minipage}[t]{0.5\linewidth}
\centering
\includegraphics[width=\linewidth]{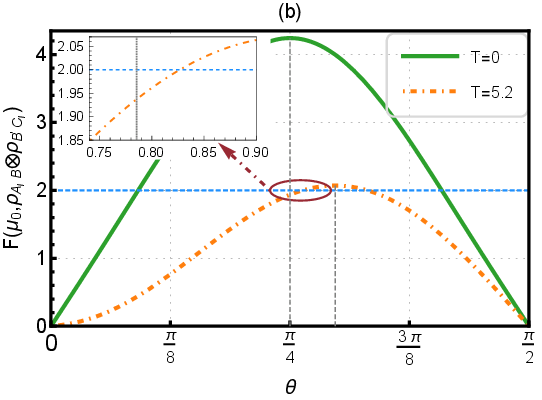}
\end{minipage}%
\caption{Network steering functional $F$ versus initial-state parameter $\theta$ at $T=0$ and $T=5.2$, with $\omega=1$. (a) Single-terminal configuration. (b) Double-terminal configuration. The horizontal line indicates the NLHS bound $F=2$, and the vertical line marks the maximally entangled input, $\theta=\pi/4$. The inset in (b) enlarges the finite-temperature curve near the detection bound, showing that some nonmaximally entangled inputs retain a violation when the maximally entangled input does not.}
\label{Fig3}
\end{figure}
Fig.~\ref{Fig3} shows the network steering functional $F$ as a function of the initial-state parameter $\theta$ for different Hawking temperatures $T$, with $\omega=1$. In both configurations, $F$ increases from zero to a maximum and then decreases to zero as $\theta$ varies from $0$ to $\pi/2$. Network steering is certified only within the parameter intervals where $F>2$. At $T=0$, the maximum occurs at $\theta=\pi/4$, corresponding to two maximally entangled input pairs. At finite temperature, the curves are suppressed and the maximum shifts toward larger $\theta$, indicating that the initial-state parameter yielding the strongest inequality violation depends on the Hawking temperature. This shift does not imply stronger initial entanglement, since the entanglement of each input pair decreases as $\theta$ increases beyond $\pi/4$. Comparing Figs.~\ref{Fig3}(a) and \ref{Fig3}(b), the double-terminal configuration exhibits a more pronounced reduction of $F$ and a narrower interval of detectable steering. In particular, the inset of Fig.~\ref{Fig3}(b) shows that at $T=5.2$, the maximally entangled input yields $F\leq2$, whereas suitably chosen nonmaximally entangled inputs still yield $F>2$. Thus, maximal initial entanglement does not necessarily provide the greatest robustness for this network steering criterion. These results highlight the joint influence of the initial-state parameter and the distribution of affected nodes on the persistence of certifiable network steering under Hawking effects.

%------------------------------------------------------------------------------------------------------------------------------------------------------------------------------------------------%

%------------------------------------------------------------------------------------------------------------------------------------------------------------------------------------------------%
\section{Conclusions\label{sec:conclusions}}

We have investigated network quantum steering in a three-node linear network supplied by two independent sources of entangled massless Dirac-field qubits in Schwarzschild spacetime. By comparing configurations in which one or both end nodes are affected by Hawking effects, we examined the roles of node configuration and initial-state preparation using a network steering inequality. Hawking effects suppress the inequality violation, with the double-terminal configuration exhibiting lower robustness than the single-terminal configuration. The results further show that the initial-state parameter yielding the largest violation shifts away from the maximally entangled value at finite temperature. In particular, nonmaximally entangled inputs can retain a violation in a temperature range where maximally entangled inputs are no longer certified by the same criterion. Thus, the amount of initial entanglement alone does not determine the persistence of certifiable network steering; its response to the Hawking-induced transformation is also essential.

These findings identify initial-state preparation and node configuration as complementary considerations for preserving certifiable quantum correlations in relativistic networks. For experimental platforms where nonmaximally entangled states are easier to prepare, their favorable performance may also offer a practical advantage. The conclusions apply to the field-mode approximation and measurement configuration adopted here. Moreover, the inequality provides a sufficient criterion for network steering: a value $F\leq2$ indicates that steering is not detected by this criterion, rather than establishing the existence of an NLHS model. Future work could optimize the joint measurement and explore more general field-mode descriptions and network structures to determine how broadly the observed advantages persist.

\begin{acknowledgments}
This work is supported by the National Natural Science Foundation of China (Grant Nos. 12675070 and 12575056), the Natural
Science Foundation of the Liaoning Scientific Committee (Grant No. 2026-MS-281), and Liaoning Revitalization Talents Program (XLYC2007047 and XLYC2503099).
\end{acknowledgments}
\section*{Data availability}
The data supporting the findings of this study are available from the corresponding author, Shu-Min Wu, upon request.


\begin{thebibliography}{99}
\bibitem{M18}
E. Schr\"{o}dinger, Discussion of Probability Relations between Separated Systems, Math. Proc. Cambridge Philos. Soc. {\bf31}, 555 (1935).
\bibitem{M19}
A. Einstein, B. Podolsky, and N. Rosen, Can Quantum-Mechanical Description of Physical Reality Be Considered Complete?, Phys. Rev. {\bf47}, 777 (1935).

\bibitem{M24}
S. J. Jones, H. M. Wiseman, and A. C. Doherty, Entanglement, Einstein-Podolsky-Rosen correlations, Bell nonlocality, and steering, Phys. Rev. A {\bf76}, 052116 (2007).
\bibitem{M25}
D. J. Saunders, S. J. Jones, H. M. Wiseman, and G. J. Pryde, Experimental EPR-steering using Bell-local states, Nat. Phys. {\bf6}, 845 (2010).

\bibitem{M20}
H. M. Wiseman, S. J. Jones, and A. C. Doherty, Steering, Entanglement, Nonlocality, and the EPR Paradox, Phys. Rev. Lett. {\bf98}, 140402 (2007).
\bibitem{M21}
O. G\"{u}hne and  G. T\'{o}th, Entanglement detection, Phys. Rep. {\bf474}, 1 (2009).
\bibitem{M22}
R. Horodecki, P. Horodecki, M. Horodecki, and K. Horodecki, Quantum entanglement, Rev. Mod. Phys. {\bf81}, 865 (2009).
\bibitem{M23}
N. Brunner, D. Cavalcanti, S. Pironio, V. Scarani, and S. Wehner, Bell nonlocality, Rev. Mod. Phys. {\bf86}, 419 (2014).




\bibitem{M43}
P. T\"{o}rm\"{a}, Transitions in Quantum Networks, Phys. Rev. Lett. {\bf81}, 2185 (1998).
\bibitem{M44}
S. Wehner, D. Elkouss, and R. Hanson, Quantum internet: A vision for the road ahead, Science {\bf362},  eaam9288 (2018).
\bibitem{M45}
J. \r{A}berg, R. Nery, C. Duarte, and R. Chaves, Semidefinite Tests for Quantum Network Topologies, Phys. Rev. Lett. {\bf125}, 110505 (2020).
\bibitem{M46}
C. Branciard, N. Gisin, and S. Pironio, Characterizing the Nonlocal Correlations Created via Entanglement Swapping, Phys. Rev. Lett. {\bf104}, 170401 (2010).
\bibitem{M47}
C. Branciard, D. Rosset, N. Gisin, and S. Pironio, Bilocal versus nonbilocal correlations in entanglement-swapping experiments, Phys. Rev. A {\bf85}, 032119 (2012).
\bibitem{M48}
A. Tavakoli, P. Skrzypczyk, D. Cavalcanti, and A. Ac\'{i}n, Nonlocal correlations in the star-network configuration, Phys. Rev. A {\bf90}, 062109 (2014).
\bibitem{M49}
M. O. Renou, E. B\"{a}umer, S. Boreiri, N. Brunner, N. Gisin, and S. Beigi, Genuine Quantum Nonlocality in the Triangle Network, Phys. Rev. Lett. {\bf123}, 140401 (2019).
\bibitem{M50}
S. Munshi, R. Kumar, and A. K. Pan, Generalized n-locality inequalities in a star-network configuration and their optimal quantum violations, Phys. Rev. A {\bf104}, 042217 (2021).
\bibitem{M51}
L. Yang, X. Qi, and J. Hou, Nonlocal correlations in the tree-tensor-network configuration, Phys. Rev. A {\bf104}, 042405 (2021).
\bibitem{M52}
A. Suprano, D. Poderini, E. Polino, I. Agresti, G. Carvacho, A. Canabarro, E. Wolfe, R. Chaves, and F. Sciarrino, Experimental Genuine Tripartite Nonlocality in a Quantum Triangle Network, PRX Quantum {\bf3}, 030342 (2022).
\bibitem{M53}
A. Tavakoli, A. Pozas-Kerstjens, M. X. Luo, and M. O. Renou, Bell nonlocality in networks, Rep. Prog. Phys. {\bf85}, 056001 (2022).
\bibitem{M54}
B. D. M. Jones,  I. \v{S}upi\'{c}, R. Uola, N. Brunner, and P. Skrzypczyk, Network quantum steering, Phys. Rev. Lett. {\bf127}, 170405 (2021).
\bibitem{M55}
Q. Y. He and M. D. Reid, Genuine Multipartite Einstein-Podolsky-Rosen Steering, Phys. Rev. Lett. {\bf111}, 250403 (2013).
\bibitem{M56}
R. V. Nery, M. M. Taddei, R. Chaves, and L. Aolita, Quantum Steering Beyond Instrumental Causal Networks, Phys. Rev. Lett. {\bf120}, 140408 (2018).
\bibitem{M57}
D. Cavalcanti, P. Skrzypczyk, G. H. Aguilar, R. V. Nery, P. H. S. Ribeiro, and S. P. Walborn, Detection of entanglement in asymmetric quantum networks and multipartite quantum steering, Nat. Commun. {\bf6}, 7941 (2015).
\bibitem{M58}
A. Tavakoli, N. Gisin, and C. Branciard, Bilocal Bell Inequalities Violated by the Quantum Elegant Joint Measurement, Phys. Rev. Lett. {\bf126}, 220401 (2021).
\bibitem{M60}
Y. Xiang, S. Cheng, Q. Gong, Z. Ficek, and Q. He, Quantum Steering: Practical Challenges and Future Directions, PRX Quantum {\bf3}, 030102 (2022).
\bibitem{M61}
Z. Chen, K. Wu, and S. M. Fei, Detection of network and genuine network quantum steering, Results Phys. {\bf56}, 107244 (2024).
\bibitem{M62}
G. Jiang, X. Wu, and T. Zhou, Quantum steering in a star network, Europhys. Lett. {\bf146}, 68001 (2024).
\bibitem{M63}
Y. Li, Y. Xiang, X. D. Yu, H. C. Nguyen, O. G\"{u}hne, and Q. He, Randomness Certification from Multipartite Quantum Steering for Arbitrary Dimensional Systems, Phys. Rev. Lett. {\bf132}, 080201 (2024).





\bibitem{SDF1}
W. Liu, C. Wen, and J. Wang, Lorentz violation alleviates gravitationally induced entanglement degradation, J. High Energy Phys. {\bf2025}, 184 (2025).

\bibitem{SDF2}
S. Sen, A. Mukherjee, and S. Gangopadhyay, Entanglement degradation as a tool to detect signatures of modified gravity, Phys. Rev. D {\bf109}, 046012 (2024).

\bibitem{SDF3}
I. Fuentes-Schuller and  R. B. Mann, Alice falls into a black hole: Entanglement in noninertial frames, Phys. Rev. Lett. {\bf95}, 120404 (2005).

\bibitem{SDF4}
Q. Pan and J. Jing, Hawking radiation, entanglement, and teleportation in the background of an asymptotically flat static black hole, Phys. Rev. D {\bf78}, 065015 (2008).

\bibitem{SDF5}
H. Wu and L. Chen, Orbital angular momentum entanglement in noninertial reference frame, Phys. Rev. D {\bf107}, 065006 (2023).

\bibitem{SDF6}
I. Agullo, A. Delhom, and \'A. Parra-L\'opez, Toward the observation of entangled pairs in BEC analog expanding universes, Phys. Rev. D \textbf{110}, 125023 (2024).


\bibitem{SDF7}
E. Mart\'{\i}n-Mart\'{\i}nez, L. J. Garay, and J. Le\'{o}n, Unveiling quantum entanglement degradation near a Schwarzschild black hole,  Phys. Rev. D {\bf82}, 064006 (2010).

\bibitem{SDF8}
P. M. Alsing, I. Fuentes-Schuller, R. B. Mann, and T. E. Tessier, Entanglement of Dirac fields in noninertial frames, Phys. Rev. A {\bf74}, 032326 (2006).

\bibitem{SDF9}
X. Liu, C. Zeng, and J. Wang, Generation of quantum entanglement in superposed diamond spacetime, Eur. Phys. J. C {\bf85}, 539 (2025).


\bibitem{SDF10}
M. M. Du, H. W. Li, S. T. Shen, X. J. Yan, X. Y. Li, L. Zhou, W. Zhong, and Y. B. Sheng, Maximal steered coherence in the background of Schwarzschild space-time, Eur. Phys. J. C {\bf84}, 450 (2024).

\bibitem{SDF11}
S. Elghaayda, A. Ali, S. Al-Kuwari, and M. Mansour, Physically accessible and inaccessible quantum correlations of Dirac fields in Schwarzschild spacetime, Phys. Lett. A {\bf525}, 129915 (2024).


\bibitem{SDF12}
J. K. Basak, D. Giataganas, S. Mondal, and W. Y. Wen, Reflected entropy and Markov gap in noninertial frames, Phys. Rev. D {\bf108}, 125009 (2023).

\bibitem{SDF13}
Q. Xiao, Y. Chen, and T. Liu, Effects of Lorentz symmetry breaking on quantum coherence in an expanding universe,
Phys. Lett. B {\bf872}, 140133  (2026).

\bibitem{SDF14}
C. Y. Liu, Z. W. Long, and Q. L. He, Quantum coherence and quantum Fisher information of Dirac particles in curved spacetime under decoherence, Phys. Lett. B {\bf857}, 138991 (2024).


\bibitem{SDF15}
A. A. Svidzinsky, M. O. Scully, and W. Unruh, Minkowski vacuum entanglement and accelerated oscillator chains, Phys. Rev. D {\bf111}, 045022 (2025).


\bibitem{SDF16}
S. M. Wu, H. Y. Wu, Y. X. Wang, and J. Wang, Gaussian tripartite steering in Schwarzschild black hole, Phys. Lett. B {\bf865},  139493 (2025).

\bibitem{SDF17}
G. Adesso, I. Fuentes-Schuller, and M. Ericsson, Continuous-variable entanglement sharing in noninertial frames, Phys. Rev. A {\bf76}, 062112 (2007).


\bibitem{SDF18}
T. Gonzalez-Raya, S. Pirandola, and M. Sanz, Satellite-based entanglement distribution and quantum teleportation with continuous variables, Commun. Phys. {\bf7},  126 (2024).

\bibitem{SDF19}
G. Ciliberto, S. Emig, N. Pavloff, and M. Isoard, Violation of Bell inequalities in an analog black hole, Phys. Rev. A {\bf109}, 063325 (2024).

\bibitem{SDF20}
S. M. Wu and H. S. Zeng, Genuine tripartite nonlocality and entanglement in curved spacetime, Eur. Phys. J. C {\bf82}, 4 (2022).


\bibitem{AAASDF59}
Q. Liu, T. Liu, C. Wen, and J. Wang, Optimal quantum strategy for locating Unruh channels, Phys. Rev. A {\bf110}, 022428 (2024).


\bibitem{SDF21}
H. Du and R. B. Mann, Fisher information as a probe of spacetime structure: relativistic quantum metrology in (A)dS, J. High Energy Phys. {\bf2021}, 112 (2021).



\bibitem{SDF22}
S. M. Wu, X. W. Fan, R. D. Wang, H. Y. Wu, X. L. Huang, and H. S. Zeng, Does Hawking effect always degrade fidelity of quantum teleportation in Schwarzschild spacetime?, J. High Energy Phys. {\bf2023}, 232 (2023).

\bibitem{SDF23}
S. Bellucci, V. Kh. Kotanjyan, and A. A. Saharian, Fermionic condensate and the mean energy-momentum tensor
in the Fulling-Rindler vacuum, Phys. Rev. D {\bf108}, 085014 (2023).

\bibitem{SDF24}
T. Y. Wang and D. Wang, Entropic uncertainty relations in Schwarzschild space-time, Phys. Lett. B {\bf855}, 138876 (2024).

\bibitem{SDF25}
H. E. Camblong, A. Chakraborty, P. Lopez-Duque, and C. R. Ord\'{o}\~{n}ez, Entanglement degradation in causal diamonds, Phys. Rev. D {\bf109}, 105003 (2024).

\bibitem{SDF26}
B. Yu, X. Y. Yang, X. Hu, Z. X. Jin, and X. Huang, Nonlocal advantage of quantum imaginarity in Schwarzschild spacetime, arXiv:2604.03633.

\bibitem{SDF27}
A. Belfiglio, O. Luongo, and S. Mancini, Quantum entanglement in cosmology, Phys. Rep. {\bf1146}, 1 (2025).


\bibitem{SDF28}
X. Liu, W. Liu, Z. Liu, and J. Wang, Harvesting correlations from BTZ black hole coupled to a Lorentz-violating vector field, J. High Energy Phys. {\bf2025}, 94 (2025).


\bibitem{SDF29}
Z. D. Wei, W. Han, Y. J. Zhang, Z. X. Man, Y. J. Xia, and H. Fan, Effect of the gravitational redshift on the precision of phase estimation, Phys. Rev. D \textbf{111},  026007 (2025).


\bibitem{SDF30}
W. Izquierdo,  J. Beltran,  and E. Arias, Enhancement of harvesting vacuum entanglement in Cosmic String Spacetime, J. High Energy Phys. {\bf2025}, 49 (2025).


\bibitem{SDF31}
Z. Liu, W. Liu, X. Liu, and J. Wang, Wormhole-Induced correlation: A Link Between Two Universes, Phys. Lett. B {\bf879}, 140667 (2026).

\bibitem{SDF32}
S. M. Wu, C. X. Wang, D. D. Liu, X. L. Huang, and H. S. Zeng, Would quantum coherence be increased by curvature effect in de Sitter space?, J. High Energy Phys. {\bf2023}, 115  (2023).

\bibitem{SDF33}
A. Chakraborty,  L. Hackl, and M. Zych, Entanglement harvesting in quantum superposed spacetime, Phys. Rev. D {\bf111}, 104052 (2025).

\bibitem{SDF34}
Y. Tang, W. Liu, Z. Liu, and J. Wang, Can the signatures of quantum superposition be detected
through correlation harvesting?, J. High Energy Phys. {\bf2026}, 45 (2026).

\bibitem{SDF35}
Y. Tang, W. Liu, and J. Wang, Observational signature of Lorentz violation in acceleration radiation, Eur.
Phys. J. C {\bf 85}, 1108 (2025).


\bibitem{SDF36}
Z. Liu, Y. Li, Z. Tian, and J. Wang, Scrambling-Enhanced Quantum Battery Charging in Black Hole Analogues, Adv. Sci. {\bf 13}, e20281 (2026).

\bibitem{SDF37}
M. M. Du, H. W. Li, Z. Tao, S. T. Shen, X. J. Yan, X. Y. Li, W. Zhong, Y. B. Sheng, and L. Zhou, Basis-independent quantum coherence and its distribution under relativistic motion, Eur. Phys. J. C {\bf84}, 838 (2024).

\bibitem{SDF38}
S. Barman, I. Chakraborty, and S. Mukherjee, Signatures of gravitational wave memory in the radiative process of entangled quantum probes, Phys. Rev. D {\bf111}, 025021 (2025).


\bibitem{SDF39}
J. Foo,  R. B. Mann, and M. Zych, Entanglement amplification between superposed detectors in flat and curved spacetimes,
Phys. Rev. D {\bf103}, 065013  (2021).


\bibitem{SDF40}
S. Li and X. Huang, Geometric quantum discord in the black hole quantum atmosphere, Eur. Phys. J. C {\bf86}, 1107 (2026).

\bibitem{SDF41}
A. Zhang, Y. Zheng, X. Huang, and T. Zhang, Quantum correlations of tripartite mixed states in the black hole quantum atmosphere, Eur. Phys. J. C  {\bf86}, 1105 (2026).


\bibitem{SDF42}
J. Foo, C. S. Arabaci, M. Zych, and R. B. Mann, Quantum Signatures of Black Hole Mass Superpositions, Phys. Rev. Lett. {\bf129}, 181301  (2022).


\bibitem{SDF43}
R. Sun and X. Huang, Ergotropy of multipartite entangled states in the quantum atmosphere, arXiv:2610.18028.

\bibitem{SD1}
Y. K. Zhang, T. Aziz, L. J. Li, X. K. Song, L. Ye, and D. Wang, Quantumness and entropic uncertainty for a pair of static Unruh-DeWitt detectors, Phys. Lett. B {\bf873}, 140167 (2026).


\bibitem{QGL1}
W. Liu, D. Wu, and J. Wang, Light rings and shadows of static black holes in effective quantum gravity II: A new solution without Cauchy horizons, Phys. Lett. B {\bf868}, 139742 (2025).

\bibitem{QGL2}
W. Liu, D. Wu, and J. Wang, Light rings and shadows of static black holes in effective quantum gravity, Phys. Lett. B {\bf858}, 139052 (2024).


\bibitem{SD2}
L. J. Li, X. K. Song, L. Ye, and D. Wang, Quantifying quantumness in (A)dS spacetimes with Unruh-DeWitt detector, Phys. Rev. D {\bf111}, 065007 (2025).


\bibitem{QGL3}
B. Yu, X. Y. Yang, X. Hu, Z. X. Jin, and X. Huang, Quantum Steering of Imaginarity in Schwarzschild Spacetime, Ann. Phys. (Berlin) {\bf538}, e70267 (2026).


\bibitem{QGL4}
S. Li, Y. Yuan, and X. Huang, Performance of quantum imaginarity in black hole spacetime, arXiv:2609.00573.


\bibitem{SD3}
Y. K. Zhang, L. J. Li, X. K. Song, L. Ye, and D. Wang, Entropic uncertainty and quantum non-classicality of Unruh-Dewitt detectors in relativity, Phys. Lett. B {\bf858}, 139063 (2024).


\bibitem{QGL5}
J. Foo, S. Onoe, M. Zych, and T. C. Ralph, Generating multi-partite entanglement from the quantum vacuum with a finite-lifetime mirror, New J. Phys. {\bf22}, 083075  (2020).

\bibitem{QGL6}
J. Foo and M. Zych, Superpositions of thermalisations in relativistic quantum field theory, Quantum {\bf9}, 1629 (2025).

\bibitem{QGL7}
S. Barman and K. Hari, Radiative process of tripartite entangled probes in inertial motion, Phys. Rev. D {\bf113}, 045012 (2026).


\bibitem{QGL8}
Mayank, K. Hari, S. Barman, and D. Kothawala, Entanglement between accelerated probes in a de Sitter spacetime, Phys. Rev. D {\bf111}, 105022 (2025).

\bibitem{QGL9}
S. H. Li, T. Y.  Wang, A. Y. Tong, and S. M. Wu, Nonlocal correlation in quantum network under relativistic motion,  arXiv:2608.20650.

\bibitem{QGL10}
X. Liu, Z. Tian, and J. Wang, Enhancement of Unruh quantum Otto engine by quadratic coupling,
Sci. China Phys. Mech. Astron. {\bf69},  100412  (2026).


\bibitem{SDF45}
S. H. Li, S. H. Shang, and S. M. Wu, Does acceleration always degrade quantum entanglement for tetrapartite Unruh-DeWitt detectors?, J. High Energy Phys. {\bf2025}, 214  (2025).


\bibitem{SDF46}
S. Harikrishnan, S. Jambulingam, P. P. Rohde, and C. Radhakrishnan, Accessible and inaccessible quantum coherence in relativistic quantum systems, Phys. Rev. A {\bf105}, 052403 (2022).

\bibitem{SDF47}
G. W. Mi, X. Huang, S. M. Fei, and T. Zhang, Genuine four-partite Bell nonlocality in the curved spacetime, Eur. Phys. J. C {\bf85}, 354 (2025).


\bibitem{SDF48}
A. J. Torres-Arenas, Q. Dong, G. H. Sun, W. C. Qiang, and S. H. Dong, Entanglement measures of W-state in noninertial frames, Phys. Lett. B {\bf789}, 93 (2019).


\bibitem{SDF49}
W. M. Li and S. M. Wu, Bosonic and fermionic coherence of N-partite states in the background of a dilaton black hole, J. High Energy Phys. {\bf2024}, 144 (2024).


\bibitem{SDF50}
S. Wang and X. Huang, Quantumness of Entanglement in Schwarzschild de Sitter Spacetime, arXiv:2610.22761.


\bibitem{SDF51}
S. M.  Wu, X. W. Teng, W. M. Li, Y. X. Wang, and J. Lu, Nonseparability of multipartite systems in dilaton black hole, JCAP {\bf09}, 030  (2025).


\bibitem{SDF52}
S. M. Wu, X. W. Teng, J. X. Li, S. H. Li, T. H. Liu, and J. C. Wang, Genuinely accessible and inaccessible
entanglement in Schwarzschild black hole, Phys. Lett. B {\bf848}, 138334 (2024).


\bibitem{SDF53}
A. Ali, S. Al-Kuwari, M. Ghominejad, M. T. Rahim, D. Wang, and S. Haddadi, Quantum characteristics near event horizons, Phys. Rev. D {\bf110}, 064001 (2024).


\bibitem{MM}
M. X. Li,  Y. Li, Y. Xi, C. Y. Zhang,  Y. Z. Wang, R. B. Xu, S. M. Fei, and Z. J. Zheng, Detecting quantum steering in networks, Phys. Rev. A 111, 012624 (2025).


\bibitem{SDF54}
S. W. Hawking, Black hole explosions?, Nature {\bf248}, 30 (1974).

\bibitem{WuZeng2022}
R. J. Yao and D. Wang, Generalized entropic uncertainty relation and nonclassicality for Schwarzschild black holes, Phys. Rev. D {\bf113}, 065002 (2026).


 \bibitem{AAAA3}
D. Garfinkle, G. T. Horowitz, and A. Strominger, Charged black holes in string theory, Phys. Rev. D {\bf45}, 3888 (1992).








































\end{thebibliography}
\end{document}